\documentclass[11pt]{article}
\usepackage{moriond}

\def\mco{\multicolumn}

\def\ra{\rightarrow}

\def\be{\begin{equation}}
\def\ee{\end{equation}}
\def\bea{\begin{eqnarray}}
\def\eea{\end{eqnarray}}

\newcommand{\Photo}{}

\begin{document}
\vspace*{4cm}
\title{Top quark Physics at the LHC}

\author{Carlo Battilana, on behalf of the ATLAS and CMS collaborations}

\address{CERN, CH-1211 Gen\`eve 23, Switzerland  \\- or  -\\
CIEMAT, Av. Complutense 40,  28040, Madrid, Spain}

\maketitle\abstracts{
An overview of recent results on top quark properties and interactions is given, obtained using data
collected with the CMS and ATLAS experiments during the years 2011 and 2012 at 7 TeV and 8 TeV 
centre-of-mass energies. Measurements of top quark pair production cross sections in several top 
quark final states are reported. Moreover, cross sections for the electroweak production of single top 
quarks in both t- and tW-channels are shown. The mass of the top quark is extracted using several 
methods. Presented results also include measurements of the W helicity in top decays, the top pair 
charge asymmetry, the top quark charge and the search for anomalous couplings. Experimental 
outcomes are compared with standard model predictions and a combination of measurements 
between the different LHC experiments is reported when available.
}

%%%%%%%%%%%%%%%%%%%%%%%%%%%%%%%%%%%%%%%%%%%%%%%%%%
% INTRODUCTION
%%%%%%%%%%%%%%%%%%%%%%%%%%%%%%%%%%%%%%%%%%%%%%%%%%

\section{Introduction}
The top quark (\textit{t}) is the heaviest of the known fundamental particles.
Its huge mass, comparable with the one of a Rhenium atom ($Z = 75$), implies that it decays before
fragmenting into hadrons. This offers the unique possibility to measure the properties of a "bare"
quark without the necessity of disentangling hadronisation effects.
The large value of $m_t$ also turns into a Yukawa coupling to the Higgs boson close to unity. For 
this reason, it is often believed that the \textit{t} can play a special role in the electroweak symmetry 
breaking mechanism.
The \textit{t} quark is part of many possible beyond the Standard Model (BSM) signatures,
where it is decay mode of, yet unknown, heavy particles. It is also foreseen that new particles can
 originate from top decays.

The Large Hadron Collider \cite{intro:lhc} (LHC) has operated remarkably well in 2011 and 2012, 
providing both the ATLAS \cite{intro:atlas} and CMS \cite{intro:cms} experiments with over a million
(around ten millions) of top quarks produced at a centre-of-mass energy of 7 TeV (8 TeV).
In the following, the status of the  top quark physics programme carried out by the two collaborations 
is reported.
A wide overview is given, covering measurements related to production of tops, generated in pairs 
and singly, as well as analyses aimed at evaluating the characteristics of the quark itself (such as 
the mass) and at understanding the properties of its decay.
A complete report of the whole set of analyses on top quark physics performed  by ATLAS and CMS
can not be fully condensed in such a short article. Focus will therefore be given to most recent results,
referring interested readers to \cite{intro:atlas_results} and \cite{intro:cms_results} 
for a complete overview.

%%%%%%%%%%%%%%%%%%%%%%%%%%%%%%%%%%%%%%%%%%%%%%%%%%
% TTbar INCLUSIVE PRODUCTION
%%%%%%%%%%%%%%%%%%%%%%%%%%%%%%%%%%%%%%%%%%%%%%%%%%

\section{Inclusive production of top pairs}\label{sec:ttbar_incl_prod}

Production of top-antitop pairs ($t\bar{t}$) happens mainly via gluon-gluon fusion and quark-antiquark
annihilation. At the LHC the former mechanism dominates, contributing for a factor $\sim$80\% to the 
generation process. Due to the fact that the Cabibbo-Kobayashi-Maskawa (CKM) $|V_{tb}|$ vertex
is close to unity, tops almost uniquely decay via the $t \ra Wb$ process. Therefore, $t\bar{t}$ 
signatures can be classified according to the combinatorics of the W boson decay.
Experimental measurements of $\sigma_{t\bar{t}}$ have been performed by both experiments using
$\sqrt{s}$ = 7 TeV LHC collisions data and profiting of all possible $t\bar{t}$ decay modes: the fully 
hadronic ($46\%$) the lepton+jets ($45\%$) and the dileptonic ($9\%$) channels.

\begin{figure}
\begin{minipage}{0.55\linewidth}
\centerline{\includegraphics[width=0.9\linewidth]{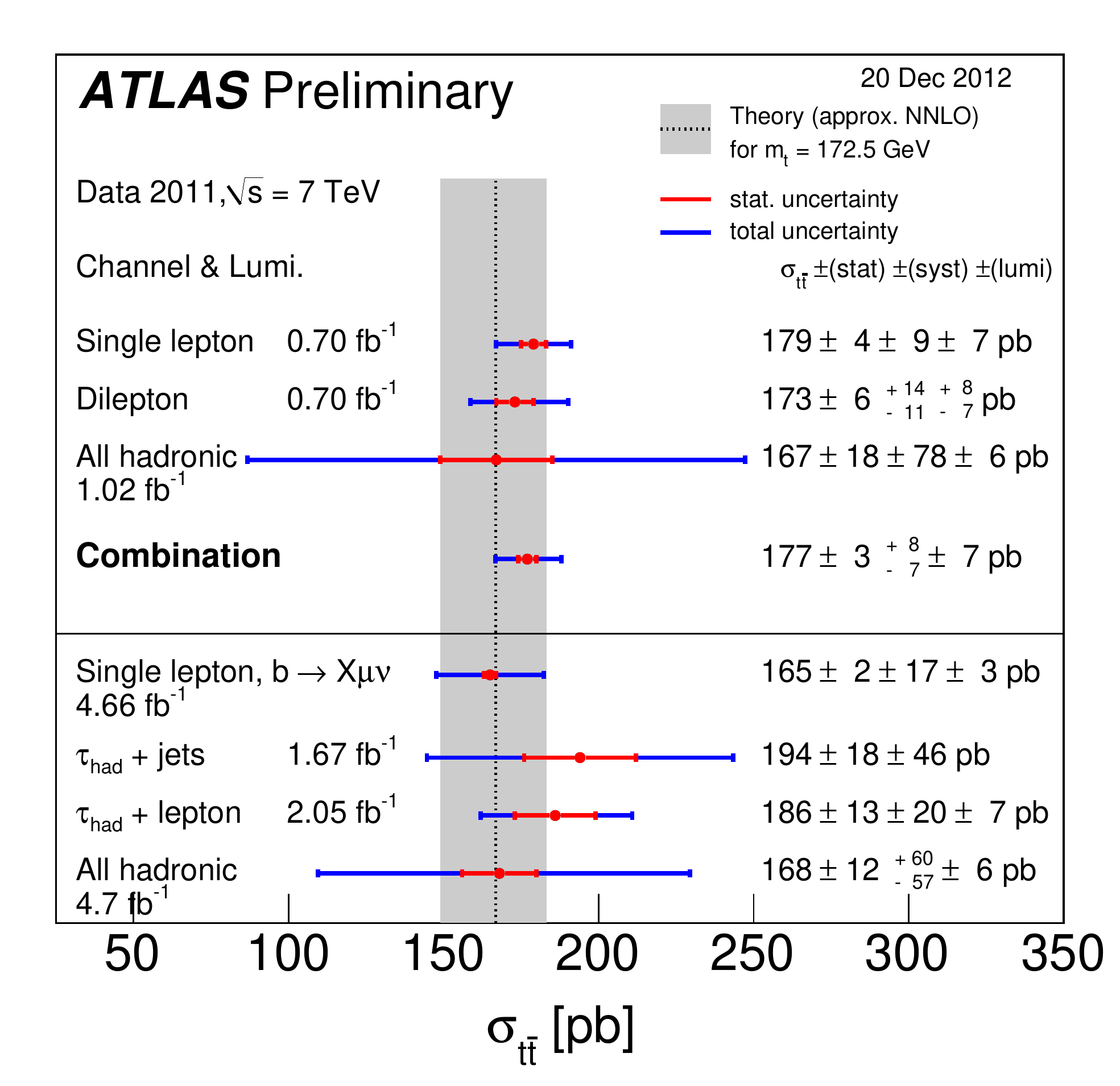}}
\end{minipage}
\hfill
\begin{minipage}{0.41\linewidth}
\centerline{\includegraphics[width=0.9\linewidth]{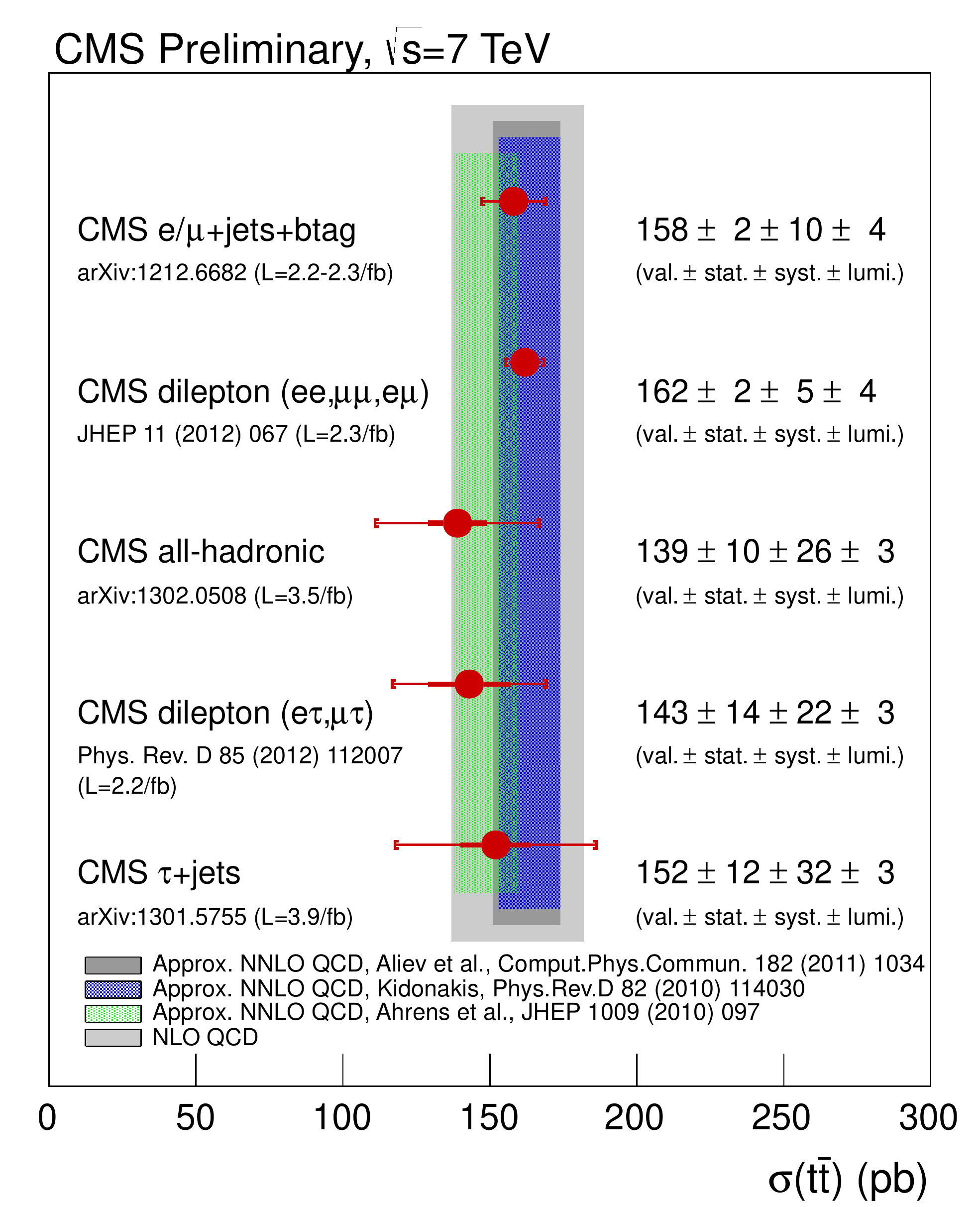}}
\end{minipage}
\hfill
\caption{Summary of ATLAS (left) and CMS (right) results on inclusive $\sigma_{t\bar{t}}$ 
	     measurements at 7 TeV computed using different decay topologies.}
\label{fig:cross_atlas_cms}
\end{figure}

For what concerns the lepton+jets channel, ATLAS reports a result of $\sigma_{t\bar{t}} = 179.0$  
$\pm 9.8$ (stat.+syst.) $\pm 6.6$ (lumi.) pb \cite{cross:atlas_lj} (precision $~7\%$) based on 0.7 
fb$^{-1}$ of data. The fraction of $t\bar{t}$ events in the sample was extracted by means of a 
likelihood fit, applied on a discriminant built by combining various kinematic variables and aimed
at distinguishing the signal from the dominant background (W+jets). 
The CMS measurement \cite{cross:cms_lj} refers to an integrated luminosity of $L$ = 2.3 fb$^{-1}$. 
A result of $\sigma_{t\bar{t}} = 158.1$  $\pm 2.1$ (stat.) $\pm 10.2$ (syst.) $\pm 3.5$ (lumi.) pb
(precision $~7\%$), was obtained by extracting the signal contribution fitting, as discriminant 
variable, the secondary vertex mass distribution.  
Both analyses have been performed in bins of jet (and b-tag) multiplicities, in order to 
constrain the contamination of background processes that are expected to have lower jet 
multiplicity (and heavy flavour content) than the signal. On top of background normalisation effects, 
major systematics uncertainties were also accounted for into the fits as nuisance parameters.
The most precise LHC measurement of $\sigma_{t\bar{t}}$ corresponds to the CMS analysis
described in \cite{cross:cms_dil}. 
It was performed in the dilepton channel using  a data sample of $L$ = 2.3 fb$^{-1}$.
In this analysis signal events are discriminated from background on the basis of the multiplicity 
distribution of jets and b-tagged jets.
The signal fraction is extracted using a fit, where systematics are accounted as nuisance parameters
into the likelihood and profiled. The reported result of $\sigma_{t\bar{t}} = 161.9$  $\pm 2.5$ (stat.) 
$^{+5.1}_{5.0}$ (syst.) $\pm 3.6$ (lumi.) pb, has relative uncertainty of $4\%$.
Other measurements, less precise, were performed by both experiments on the fully hadronic 
channel \cite{cross:altas_jets,cross:cms_jets} and channels with $\tau$
decays \cite{cross:atlas_taujets,cross:atlas_taulep,cross:cms_taujets,cross:cms_taudilep}. 
Overall results are in agreement with theoretical expectations \cite{cross:theory}.
Figure \ref{fig:cross_atlas_cms} summarises the results obtained by the two collaborations,
comparison with approximate NNLO calculations is also performed.

The scaling of  $\sigma_{t\bar{t}}$ as a function of centre-of-mass energy was also measured profiting
of 8 TeV data. CMS outcomes from analyses of $L$ = 2.8 fb$^{-1}$ performed both, in the lepton + jets
\cite{cross:cms_lj_8} and dilepton \cite{cross:cms_dil_8} channels, were combined obtaining a result
of $\sigma_{t\bar{t}} = 228$  $\pm 9$ (stat.) $\pm 27$ (syst.) $\pm 10$ (lumi.) pb. In the former case, a
strategy similar to the one already described for lepton + jets at 7 TeV was adopted, while the result 
for the latter comes from a robust cut-and-count analysis.
ATLAS  measurement is based on the study of $L$ = 5.8 fb$^{-1}$ of data in the lepton + jets channel
\cite{cross:altas_lj_8}. Signal is discriminated from background using a multivariate technique.
The final result is $\sigma_{t\bar{t}} = 241$  $\pm 2$ (stat.) $\pm 31$ (syst.) $\pm 9$ (lumi.) pb.
The various measurements from ATLAS as a function of $\sqrt{s}$ are shown in Figure 
\ref{fig:cross_scaling_diff} (left), where a comparison with approximate NNLO theoretical calculation 
is also given.

%%%%%%%%%%%%%%%%%%%%%%%%%%%%%%%%%%%%%%%%%%%%%%%%%%
% TTbar DIFFERENTIAL ANALYSES
%%%%%%%%%%%%%%%%%%%%%%%%%%%%%%%%%%%%%%%%%%%%%%%%%%

\section{Differential measurements}\label{sec:ttbar_diff_prod}

Besides inclusive measurement of $\sigma_{t\bar{t}}$, the large sample of tops collected at the LHC
allowed to perform detailed differential studies. Normalised differential cross section, computed as a 
function of relevant parameters, such as $M_{t\bar{t}}$, $y_{t\bar{t}}$, $p_{T}^{t\bar{t}}$, $p_{T}^{top}$ 
were compared with theoretical calculations, as well as with different MC models (e.g. MC@NLO,
POWHEG, MadGraph, ALPGEN).
These measurements are very precise tests of perturbative QCD, moreover they are fundamental for 
other Standard Model (SM) analyses (Higgs) and studies of BSM processes, where the top quark is 
either a dominant background, or part of the signature of a new physics signal.

\begin{figure}
\begin{minipage}{0.57\linewidth}
\centerline{\includegraphics[width=0.9\linewidth]{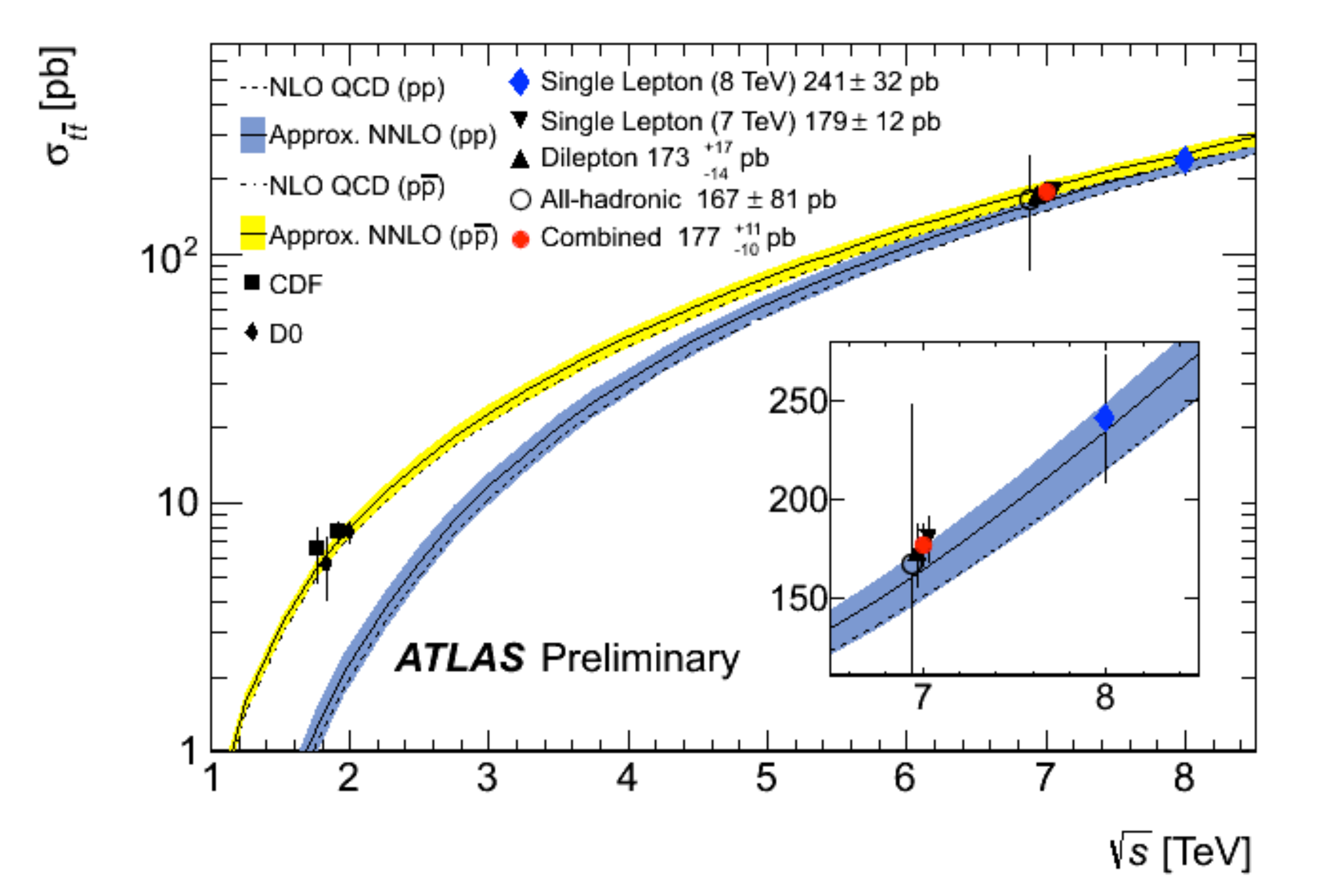}}

\end{minipage}
\hfill
\begin{minipage}{0.43\linewidth}
\centerline{\includegraphics[width=0.9\linewidth]{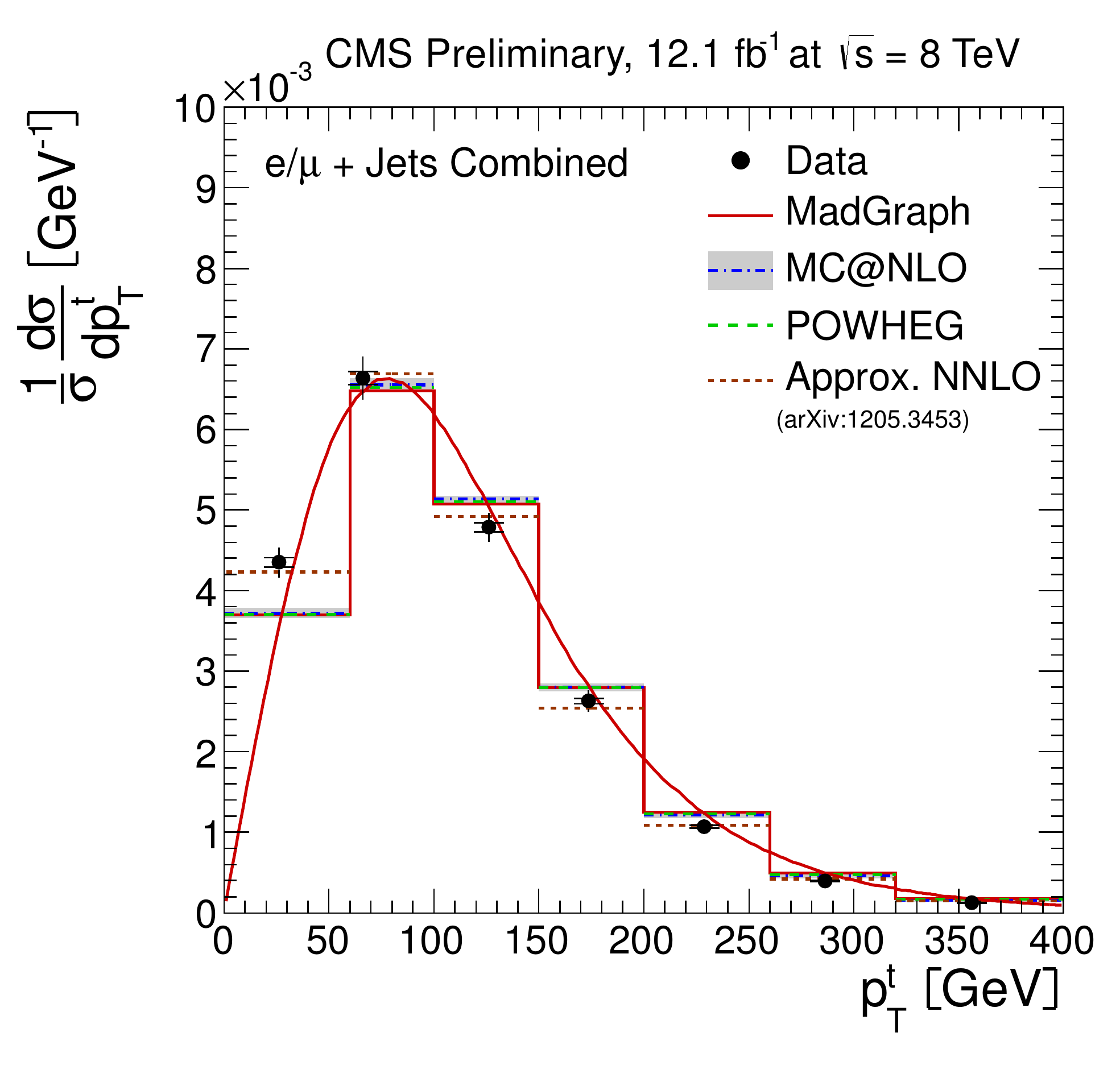}}
\end{minipage}
\hfill
\caption{Summary of ATLAS results on $\sigma_{t\bar{t}}$ showing the scaling as a function of
              $\sqrt{s}$ and its comparison with theoretical predictions (left). Differential measurement of
              normalised cross section as function of  $p_{T}^{top}$ performed by the CMS experiment in an 
              8 TeV data sample (right). Experimental results are compared with different MC models as
              well as with an approximate NNLO calculation.}
\label{fig:cross_scaling_diff}
\end{figure}

Experimentally observed quantities have been used to reconstruct the kinematics of the $t\bar{t}$
system, that has been unfolded to the level of stable hadrons, or to parton level, and extrapolated 
to the full phase space, in order to ease the comparison between experiments and with theoretical
predictions.  
Results from ATLAS are reported for the lepton + jets channel \cite{cross:altas_lj_diff} on a sample of 
$L$ = 2.03 fb$^{-1}$ of 7 TeV data. CMS has recently presented new results coming from the 
processing of $L$ = 12.1 fb$^{-1}$ of 8 TeV data both in the lepton + jets \cite{cross:cms_lj_diff} and
dilepton \cite{cross:cms_dil_diff} channels.
In case of lepton+jets analyses, the $t\bar{t}$ kinematics can be fully reconstructed by means of
kinematic fits where b-tagging information can be used to improve the results. 
In the case of the dilepton study, the presence of two neutrinos in the final state leads to an 
under-constrained kinematical system. 
The adopted strategy was then to reconstruct the $t\bar{t}$ kinematics by imposing
$m_{top}$ as an additional constraint. A scan of the results of kinematical fitting obtained by fixing
$m_{top}$ in a [100:300] $GeV/c^2$ range was performed. Results for the neutrino kinematical 
parameters obtained in this way were compared with expectations from simulation. The combination
from the scan with highest compatibility  with Monte Carlo (MC), having the larger number of b-tagged
jets, was chosen as solution.
The measurement of the normalised transverse momentum distribution of the $t$ quark performed by 
CMS in \cite{cross:cms_lj_diff} is shown in Figure \ref{fig:cross_scaling_diff} (right). While data exhibits
a softer $p_T^{top}$ dependence compared to MadGraph, MC@NLO and POWHEG, results are quite
in agreement with estimation from approximate NNLO calculations from \cite{cross:theo_diff}. All other 
results ($M_{t\bar{t}}$, $y_{t\bar{t}}$, $p_{T}^{t\bar{t}}$) have been found to be well compatible with 
current MC models.

Jet multiplicity in $t\bar{t}$ production was also measured. Additional jets can be generated together 
with top anti-top pairs through higher order QCD diagrams and their presence affects the $p_{T}$
spectra of the $t\bar{t}$ system. A direct measurement of jet multiplicity provides thus important 
information to asses how well this effect is described by initial state radiation models in simulation.
On top of that, $t\bar{t}$+jets is an important background for $ttH$ associated production and for
many BSM signals, thus its understanding is crucial for these kind of studies.

Results from both collaborations have been published profiting from the full 7 TeV data sample
\cite{cross:atlas_lj_jets,cross:cms_lj_jets,cross:cms_dil_jets}.
Figure \ref{fig:cross_jet_asymm} (left) compares jet multiplicity distributions from ATLAS ($\mu$+jets
channel) with different MC models. Results indicate that, though uncertainties are still large, MC@NLO
underestimates large jet multiplicities, whereas ALPGEN and POWHEG reproduce better the data. 
Similar conclusions \cite{cross:cms_lj_jets,cross:cms_dil_jets} were obtained on the basis of
measurements from the CMS Collaboration.

%%%%%%%%%%%%%%%%%%%%%%%%%%%%%%%%%%%%%%%%%%%%%%%%%%
% TTbar ASYMMETRY
%%%%%%%%%%%%%%%%%%%%%%%%%%%%%%%%%%%%%%%%%%%%%%%%%%

\section{Asymmetry in production of $t\bar{t}$ pairs}\label{sec:ttbar_asymm_prod}

One interesting leftover from the Tevatron top quark physics programme is the presence of an excess,
with respect to SM expectations, in the forward-backward  asymmetry ($A_{FB}$) of production of 
$t\bar{t}$ pairs in $p\bar{p}$ collisions at 1.96 TeV. This might be a hint for ``new physics''.
At the LHC it is not possible to measure this effect directly due to the intrinsic symmetry of $pp$ initial 
states.
However, the SM predicts correlation between the direction of motion of (anti)tops and the (anti)quarks
from which they origin. This arises, in case of top pair production via $q\bar{q}$ annihilation, from 
interference between the amplitudes of higher order processes (Born and box diagrams) and between
diagrams with initial/final state radiation (ISR/FSR).
Taking into account that valence quarks in protons carry, on average, more momentum than virtual 
antiquarks, the two effects result into the fact that tops tend to be emitted more abundantly at high 
values of $|y|$ with respect to anti tops, that tend to be more ``central''.
It is thus possible to measure an asymmetry in $t\bar{t}$ production at LHC by defining the charge 
asymmetry as:

\begin{equation}
A_{c} = \frac{N(\Delta{|y|} ) > 0 - N(\Delta{|y|} ) < 0}{N(\Delta{|y|} ) > 0 + N(\Delta{|y|} ) < 0}
\label{eq:charge_asymm}
\end{equation}

where $\Delta{|y|} = |y_{t}| - |y_{\bar{t}}|$ is the rapidity difference between $t$ and $\bar{t}$.
The SM NLO prediction for this observable has ben computed (for LHC collisions at 7 TeV) 
to be \cite{cross:asymm_theory} $A_{c} ^{SM}$ = 0.0115 $\pm$ 0.0006.

ATLAS has measured $A_{c}$ in both in the lepton+jets \cite{cross:atlas_asymm_lj} and the 
dileptonic channel \cite{cross:atlas_asymm_lj}, using respectively $L$ = 1.04 fb$^{-1}$ and $L $ = 4.7
fb$^{-1}$ of 7 TeV data. 
The same holds for CMS, that has based its analyses \cite{cross:cms_asymm_lj,cross:cms_asymm_dil}
on a data sample of $L$ = 5.0 fb$^{-1}$, collected at 7 TeV.
The asymmetry was computed from $\Delta{|y|}$, by means of kinematical reconstruction of the $t\bar{t}$
system. The four momentum of reconstructed top quarks was unfolded to account for detector effects 
and selection biases. In the case of CMS, this was achieved by means of a regularised unfolding procedure, 
based on generalised matrix inversion, where reconstruction and selection effects were accounted for separately.
Similarly ATLAS has performed unfolding, for the lepton+jets channel study, but has used response 
matrices that accounted, at the same time, for reconstruction and selection efficiency. The ATLAS dileptonic
analysis, instead, has computed $A_{c}$ at the reconstructed level and, after background subtraction,
has corrected it by a calibration curve. The latter was estimated on $t\bar{t}$ MC samples where  $A_{c}^{gen}$ 
values were changed at generated level by means of a reweighing technique.
Inclusive measurement of $A_{c}$, which have been proved to be in agreement with SM prediction, are
summarised in Table 1. In the case of ATLAS, the outcome of a combination of results from
the lepton+jets and dileptonic analyses is also reported.

\begin{table}[t]
\label{tab:asymmetry}
\vspace{0.1cm}
\begin{center}
\begin{tabular}{lcc}
\hline
Experiment & $A_{c}$ (lepton+jets) & $A_{c}$ (dileptonic) \\
\hline
\hline
\noalign{\vskip 1mm}
CMS      &   -0.004 $\pm$  0.010 (stat.) $\pm$  0.011 (syst.)    &   0.050 $\pm$  0.043 (stat.) $^{+0.010}_{+0.039}$ (syst.) \\
ATLAS   &   -0.019 $\pm$  0.028 (stat.) $\pm$  0.028 (syst.)    &   0.057 $\pm$  0.024 (stat.) $\pm$  0.015 (syst.) \\
\noalign{\vskip 1mm}
ATLAS (comb.)  &    \mco{2}{c}{0.029 $\pm$  0.018 (stat.) $\pm$  0.014 (syst.)}  \\
\noalign{\vskip 1mm}
\hline
\end{tabular}
\end{center}
\caption{Summary of experimental results on $A_{c}$ from CMS (lepton+jets, dileptonic channel) and ATLAS
(lepton+jets, dileptonic channel and their combination).}
\end{table}

\begin{figure}
\begin{minipage}{0.37\linewidth}
\centerline{\includegraphics[width=0.9\linewidth]{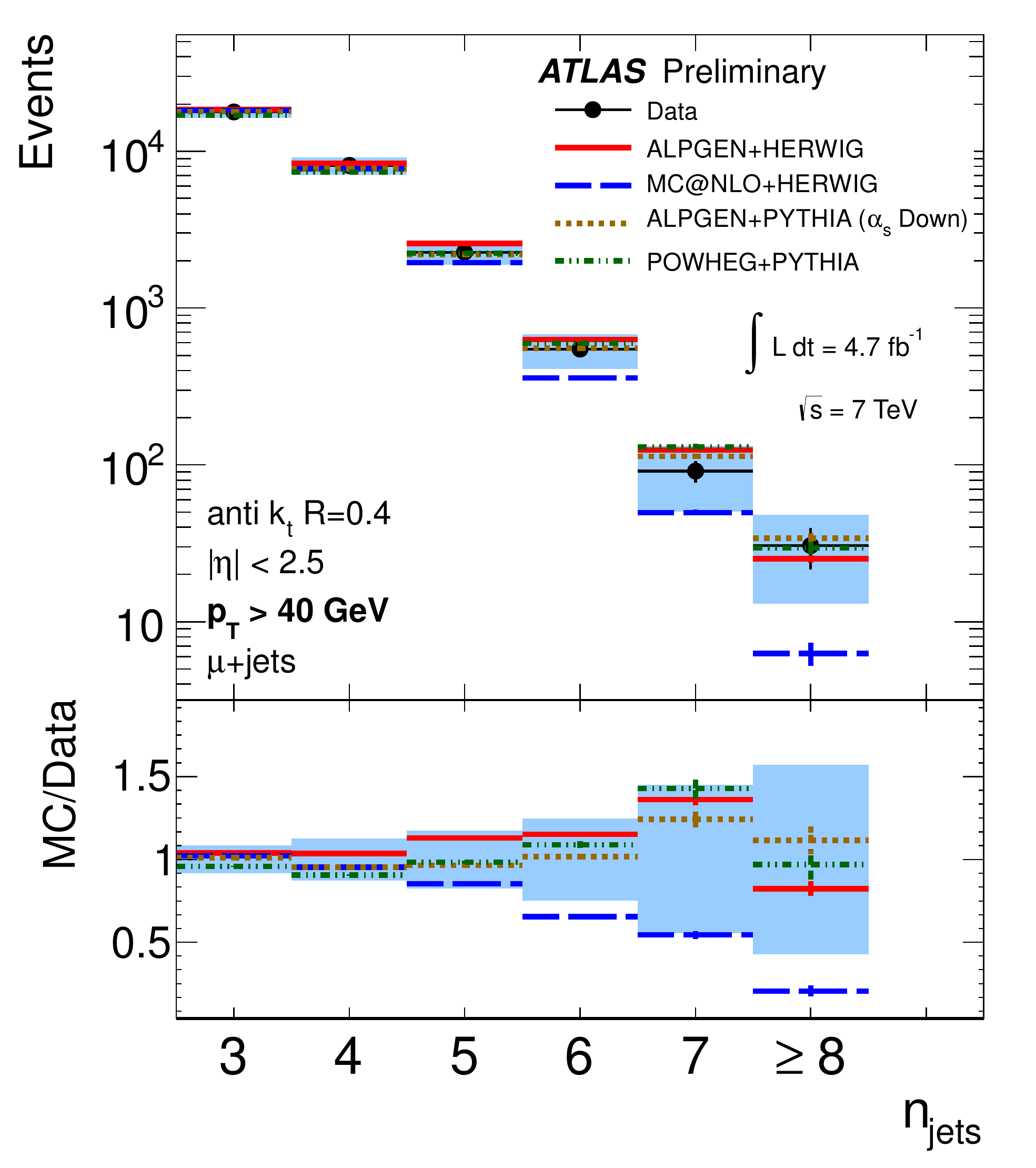}}
\end{minipage}
\hfill
\begin{minipage}{0.59\linewidth}
\centerline{\includegraphics[width=0.9\linewidth]{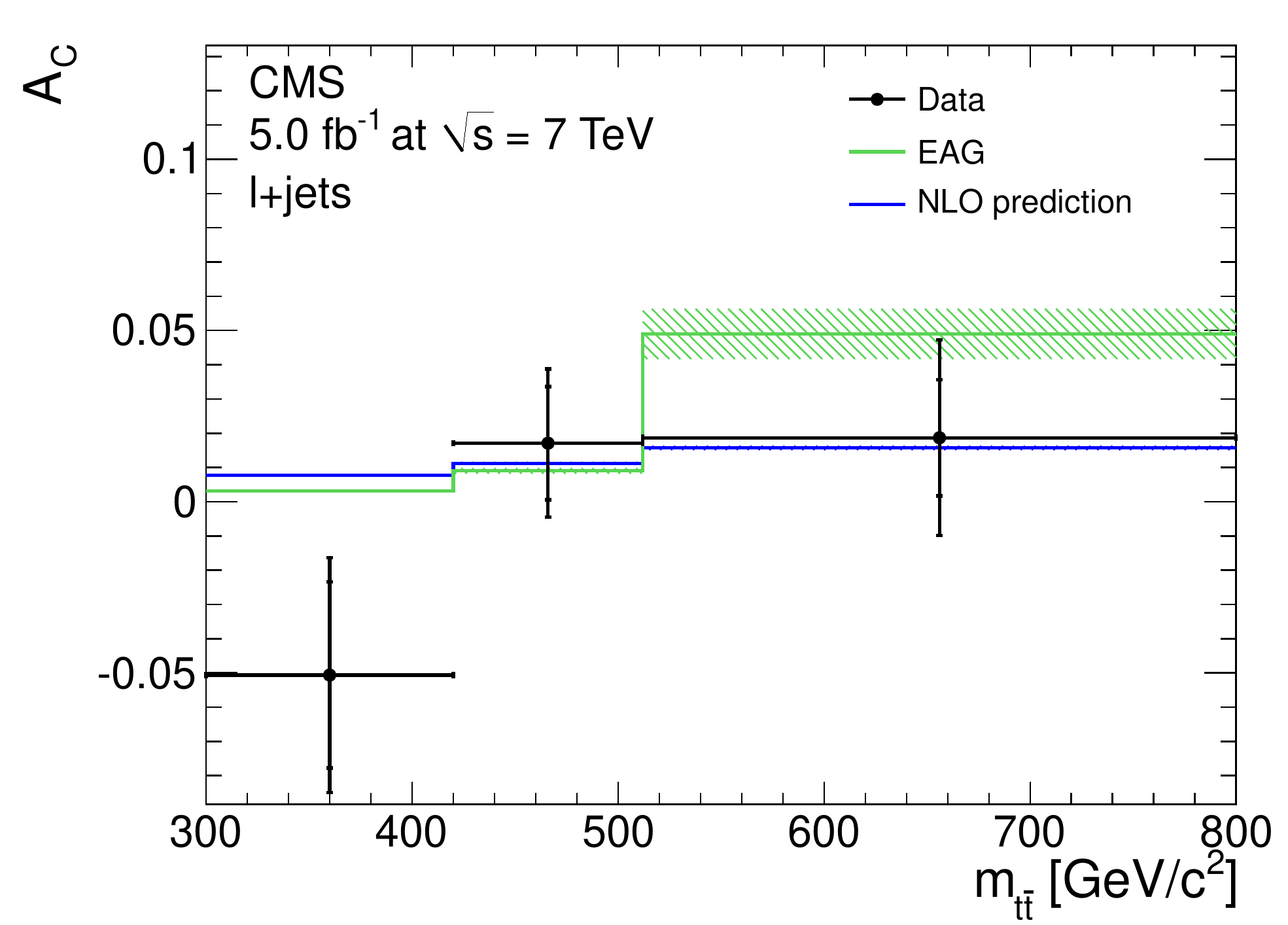}}
\end{minipage}
\hfill
\caption{Measurement of jet multiplicity in $t\bar{t}$ events in the $\mu$+jets channel by ATLAS (left).
              Results are compared with prediction from different MC generators. 
              Differential measurement of $A_{c}$ as a function of $m_{t\bar{t}}$ by CMS, compared with 
              an effective field theory model based on axial coupling of gluinos that could account for the 
              excess in $A_{FB}$ observed at Tevatron (right).}
\label{fig:cross_jet_asymm}
\end{figure}

Results have been also computed differentially and have been compared with BSM models (ATLAS) or 
effective field  theories (CMS) that might account for the $A_{FB}$ effect seen at Tevatron.
As an example, Figure \ref{fig:cross_jet_asymm} (right) depicts the differential measurement of $A_{c}$
from CMS as a function of $m_{t\bar{t}}$ and compares experimental results with the SM and an effective
field theory that could explain the excess in $A_{FB}$. The plot shows that better sensitivity needs to be
reached in order to distinguish between SM and the presented alternative model.  

%%%%%%%%%%%%%%%%%%%%%%%%%%%%%%%%%%%%%%%%%%%%%%%%%%
% Single TOP
%%%%%%%%%%%%%%%%%%%%%%%%%%%%%%%%%%%%%%%%%%%%%%%%%%

\section{Single top production}\label{sec:single_top_prod}
Besides the dominant production mechanisms that generate top pairs via strong interactions, top 
quarks are also produced singly through charge-current electroweak mechanisms.
The SM foresees the existence of three types of production channels that can generate single-tops:
the t- and s- channels, as well as the associated production of a t quark together with a W boson,
named tW-channel.
At LHC, the dominant contribution from single top production comes from the t-channel, the second 
largest one is the one from tW associated production, whereas the one for s-channel is quite small
(being ~15 times smaller than t-channel).
In all the production mechanisms, the $|V_{tb}|$ vertex is directly involved, therefore measurements
that probe single-top production cross section can be compared with theoretical predictions to 
provide information on that specific CKM matrix element.

\begin{table}[t]
\label{tab:single_top}
\vspace{0.1cm}
\begin{center}
\begin{tabular}{lccc}
\hline
Experiment & cross section (pb) & $|V_{tb}|$ unconstrained & $|V_{tb}|$ constrained \\
\hline
\hline
\noalign{\vskip 1mm}
\mco{4}{c}{t-channel (7 TeV) } \\ 
\noalign{\vskip 1mm}
\hline
ATLAS & 83 $\pm$ 4  $^{+20}_{-19}$  &  1.13 $^{+0.14}_{-0.13}$              &  $>0.75$ at $95\%$ CL \\
CMS    & 67.2 $\pm$ 6.1                      &  1.020 $\pm$ 0.046 $\pm$ 0.017 &  $>0.92$ at $95\%$ CL \\
\hline
\noalign{\vskip 1mm}
\mco{4}{c}{t-channel (8 TeV) } \\
\noalign{\vskip 1mm}
\hline
ATLAS & 95 $\pm$ 1 $\pm$ 18                            &  1.04 $^{+0.10}_{-0.11}$        &  $>0.80$ at $95\%$ CL \\
CMS    & 80.1 $\pm$ 5.7 $\pm$ 11.0 $\pm$ 4.0   &  0.96 $\pm$ 0.08 $\pm$ 0.02 &  $>0.81$ at $95\%$ CL \\
\hline
\noalign{\vskip 1mm}
\mco{4}{c}{tW-channel (7 TeV) } \\
\noalign{\vskip 1mm}
\hline
ATLAS & 16.8 $\pm$ 2.9 $\pm$ 4.9  &  1.03 $^{+0.16}_{-0.19}$                                &   \\
CMS    & 16 $^{+5}_{-4}$                  &  1.01 $^{+0.16}_{-0.13}$ $^{+0.03}_{-0.04}$  &  $>0.79$ at $90\%$ CL \\
\hline
\end{tabular}
\end{center}
\caption{Summary of experimental cross section values for single top production and
              corresponding limits on $|V_{tb}|$.}
\end{table}

Table 2 reports the status of cross sections measurements at 7 and at 8 TeV 
performed by the two collaborations, as well as the limits on $|V_{tb}|$.
Good agreement is found between data and theory \cite{singlet:theory_tch,singlet:theory_tWch}.
The description of the analyses whose results are reported in the table follows.
For what concern the measurement of the t-channel cross section at 7 TeV, CMS exploits up to 
$L$ = 1.56 fb$^{-1}$ of collected data \cite{singlet:cms_tch_7TeV}. Three analysis strategies were
carried out in parallel,  checked for compatibility, and combined to improve the final measurement.
The simplest of them computes $\sigma_{t-channel}$ on the basis of a fit on the $|\eta_{j'}|$ 
distribution of the recoiled jet from a light quark, which is expected to be boosted to high absolute 
values of rapidity in the signal signature and ensures quite good discrimination with respect to background.
On top of this very robust and simple analysis, two more complex studies, based on fits of multivariate 
discriminators coming from a neural network (NN) and a boosted decision tree (BDT), were performed. 
The same analysis strategy was also used to measure $\sigma_{t-channel}$ on $L$ = 5.0 fb$^{-1}$ of 
8 TeV data \cite{singlet:cms_tch_7TeV}.
ATLAS t-channel cross section measurements have been computed on the basis of $L$ = 1.04 
fb$^{-1}$ and $L$ = 5.8 fb$^{-1}$ of 7 and 8 TeV data 
respectively \cite{singlet:atlas_tch_7TeV,singlet:atlas_tch_8TeV}. In this case, the signal fraction was 
extracted by means of a fit performed on a NN based discriminator. 
In the case of the 7 TeV analysis, a robust cut-and-count analysis was performed in parallel and used
to validate the NN output but did not add much to the precision of the measurement itself.

One interesting observable, related to t-channel single top production in $pp$ collisions, is the 
ratio between the  cross section for tops and anti-tops. At the LHC, in fact, single t($\bar{t}$) are produced
via exchange of a virtual W boson between an u(d) and a b quark. As the PDF contribution of u and d 
quarks in protons is different, this turn into an asymmetry between $\sigma^{top}_{t-channel}$ and 
$\sigma^{anti-top}_{t-channel}$ whose ratio, at 7 TeV, is predicted to be $\sim$1.84.
This effect was measured by ATLAS and (recently) by CMS on $L$ = 4.7 fb$^{-1}$ of 7 
TeV \cite{singlet:atlas_ratio} and on $L$ = 12.2 fb$^{-1}$ of 8 TeV \cite{singlet:cms_ratio} data respectively.
Results are obtained by measuring $\sigma_{t-channel}$ independently for positive and negative 
charged tops, and making the ratio. Many systematics therefore cancel out. 
The discriminating variables used for the fits are the $|\eta_{j'}|$ distribution for CMS and the output
of a NN for ATLAS.
Results of 1.76 $\pm$ 0.27 and 1.81 $^{+0.23}_{-0.22}$ are obtained respectively by the two
experiments. Good agreement with theory is observed but an increase in the precision of the 
measurements is needed to constrain the contribution coming from different PDF sets used
in simulation.

Associated production of single tops and W bosons was also observed on 7 TeV data, firstly 
by  ATLAS \cite{singlet:atlas_tWch}  (3.3$\sigma$ significance on a $L$ = 2.05 fb$^{-1}$ sample),
then by CMS \cite{singlet:cms_tWch} (4.0$\sigma$ significance on a $L$ = 4.9 fb$^{-1}$ sample).
Both analyses extract $\sigma_{tW-channel}$ by fitting multivariate discriminators built from BDTs. 
In CMS, results are also confirmed by means of a robust cut-based study.

%%%%%%%%%%%%%%%%%%%%%%%%%%%%%%%%%%%%%%%%%%%%%%%%%%
% MASS
%%%%%%%%%%%%%%%%%%%%%%%%%%%%%%%%%%%%%%%%%%%%%%%%%%

\section{Top quark mass}\label{subsec:mass}

The mass of the $t$ quark ($m_{t}$) is a free parameter of the SM, therefore it must be experimentally
measured. It has large contribution to electroweak radiative corrections and, together with the masses 
of the $W$ and the Higgs bosons, can be used, for example, to probe the vacuum stability of the SM 
at high energy scales \cite{mass:theory_stability}.
The definition $m_{t}$ depends on the renormalisation scheme. With few exceptions, the top mass is
experimentally evaluated by computing the invariant mass of the decay products of top decays and 
calibrating the response from kinematic reconstruction with simulations. What is actually measured 
is, therefore, the value of $m_{t}$ encoded in MC generators. The latter is related with the one used 
in electroweak fits ($m_{top}^{MS}$) by $m_{top}^{MC} \simeq m_{top}^{MS} + 10$ GeV/$c^2$.
 
\begin{figure}
\begin{minipage}{0.59\linewidth}
\centerline{\includegraphics[width=0.9\linewidth]{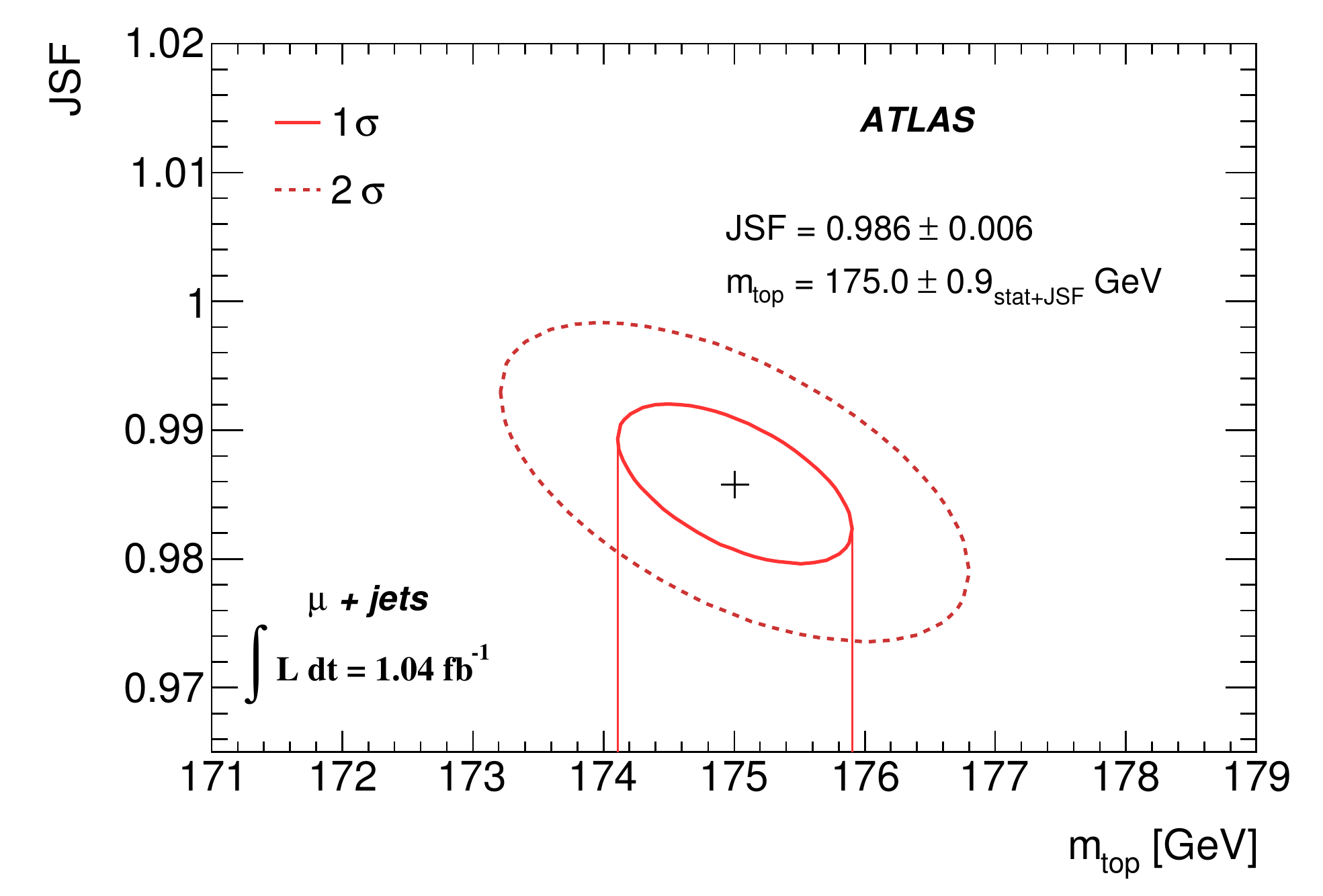}}
\end{minipage}
\hfill
\begin{minipage}{0.39\linewidth}
\centerline{\includegraphics[width=0.9\linewidth]{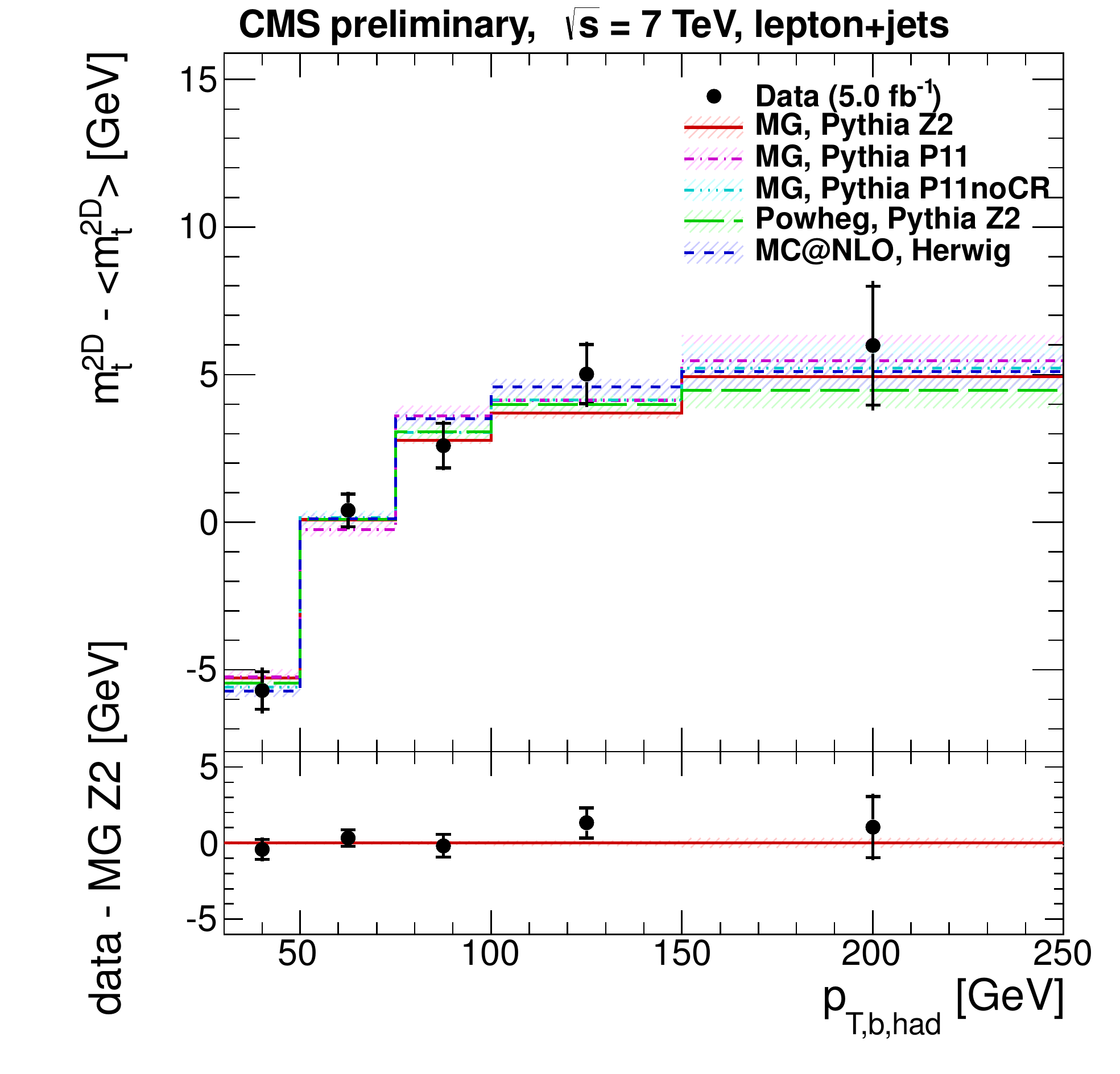}}
\end{minipage}
\hfill
\caption{The 2D likelihood used by ATLAS to extract $m_{t}$ and JSF in the $\mu$+jets final state (left). 
             Contour values of 1$\sigma$ and 2$\sigma$ are shown together with the result. 
             Differential measurement of $m_{t}$ as a function of the $p_{T}$ of the $b$-quark comparing
             data and different MC generators as well as colour reconnection tunes (right).}
\label{fig:mass_lj_diff}
\end{figure}

Experimentally, ATLAS and CMS have kinematically measured the top mass using different 
techniques and exploiting many of the $t\bar{t}$ decay channels.
As an example, measurements of $m_{t}$ in the lepton+jets channel are reported.
They were performed by ATLAS and CMS using, respectively, up to $L$ = 1.04 fb$^{-1}$ \cite{mass:atlas_lj}
and $L$ = 5.0 fb$^{-1}$ \cite{mass:cms_lj} of 7 TeV data.
In this channel, the $t\bar{t}$ system can be fully reconstructed by means of kinematic fits that constrain
the mass of the $W$ bosons and imply equality of the masses of the two decaying tops.
ATLAS measurement was performed using a template method that extracts $m_{t}$ from a likelihood
fit that compares distribution from data with different MC templates generated at different values of
$m_{t}^{gen}$. It reports a value of $m_{t}$ = 174.5 $\pm$ 0.6 (stat.) $\pm$ 2.3 (syst.) GeV/$c^2$.
In the case of CMS, the ideogram method was used. This technique exploits the full event information
and combines the results of the kinematic fit coming from all possible jet combinations by weighting 
them on the basis of a likelihood evaluated from analytic expressions from simulations.
The response of the method is calibrated using MC generated at different values of $m_{t}^{gen}$. 
A final result of $m_{t}$ = 173.49 $\pm$ 0.43 (stat+JES) $\pm$ 0.98 (syst.) GeV/$c^2$ was obtained.
This is the most precise single measurement of the top quark mass up to date.
In both cases, the largest source of systematic uncertainty comes from jet energy scale calibration.
Its effect is accounted in situ by both analyses by evaluating a scale factor (JSF) computed comparing
the reconstructed value of $m_{W}$ from the hadronic decay branch with its experimental known values.
The final measurements were performed using a fit that evaluated together $m_{t}$ and the JSF. Figure 
\ref{fig:mass_lj_diff} (left) illustrates the minimisation of the likelihood used to measure 
$m_{t}$ in the $\mu + jets$ channel of the ATLAS measurement.

As the top has colour, non-perturbative effects (known as colour reconnection) that alter the kinematics of
the top decay, hence the $m_{t}$ measurement from invariant mass of the products, exist. These effects
may depend, among others, on the kinematics of final decay state and the boost of the top quark.
Moreover, the accuracy of the  ISR/FSR modelling is a large source of systematic uncertainty for
the measurement. 
CMS has recently published a study aimed at evaluating possible biases on the kinematical measurement 
of $m_{t}$ performed using $t\bar{t}$ decay products \cite{mass:cms_diff}.
The analysis is based on $L$ = 5.0 fb$^{-1}$ of 7 TeV data in the lepton+jets channel and exploits the
ideogram method described above.
Effects such as ISR/FSR and colour reconnection were probed.
Results show good agreement between data and MC generators (MadGraph, POWHEG, MC@NLO) and
demonstrate that better sensitivity needs to be reached to distinguish between different colour 
reconnection tunes.
An example can be found in Fig. \ref{fig:mass_lj_diff} (right), that shows the differential measurement of
 $m_{t}$ as function of the $p_{T}$ of the $b$-quark (which is colour connected with the top from which
 it originates).
 
Other channels and methods were also used to measure $m_{t}$. A combination of the various 
measurements of the top mass was performed by using up to $L$ = 4.9 fb$^{-1}$ of ATLAS and CMS 7 TeV 
data \cite{mass:atlas_cms_comb}.
CMS has subsequently performed a combination of its own measurements, including more recent ones
based on up to $L$ = 5.0 fb$^{-1}$ collected at 7 TeV \cite{mass:cms_comb}.
Results are, respectively $m_{t}$ = 173.3 $\pm$ 0.5 (stat.) $\pm$ 1.3 (syst.) GeV/$c^2$ and $m_{t}$ 
= 173.36 $\pm$ 0.38 (stat.)  $\pm$ 0.91 (syst.) GeV/$c^2$.

%%%%%%%%%%%%%%%%%%%%%%%%%%%%%%%%%%%%%%%%%%%%%%%%%%
% PROPERTIES
%%%%%%%%%%%%%%%%%%%%%%%%%%%%%%%%%%%%%%%%%%%%%%%%%%

\section{Top quark properties}\label{subsec:prop}
The large statistics  of top quarks produced at the LHC has allowed to study in detail the intrinsic
properties of the particle as well as  and its decay mechanisms. Besides being a good benchmark 
of the SM, these analyses are an essential tool to check for deviations from SM that might come, 
for example, from anomalous couplings or from supersymmetric models (SUSY).
A wide set of measurements was performed, examples are the evaluation of the top charge, the
study of top polarisation and spin correlation in $t\bar{t}$ decay and many more.
In this article, focus will be given to a subset of measurements favouring the ones that have been 
recently updated.

\begin{figure}
\begin{minipage}{0.42\linewidth}
\centerline{\includegraphics[width=0.9\linewidth]{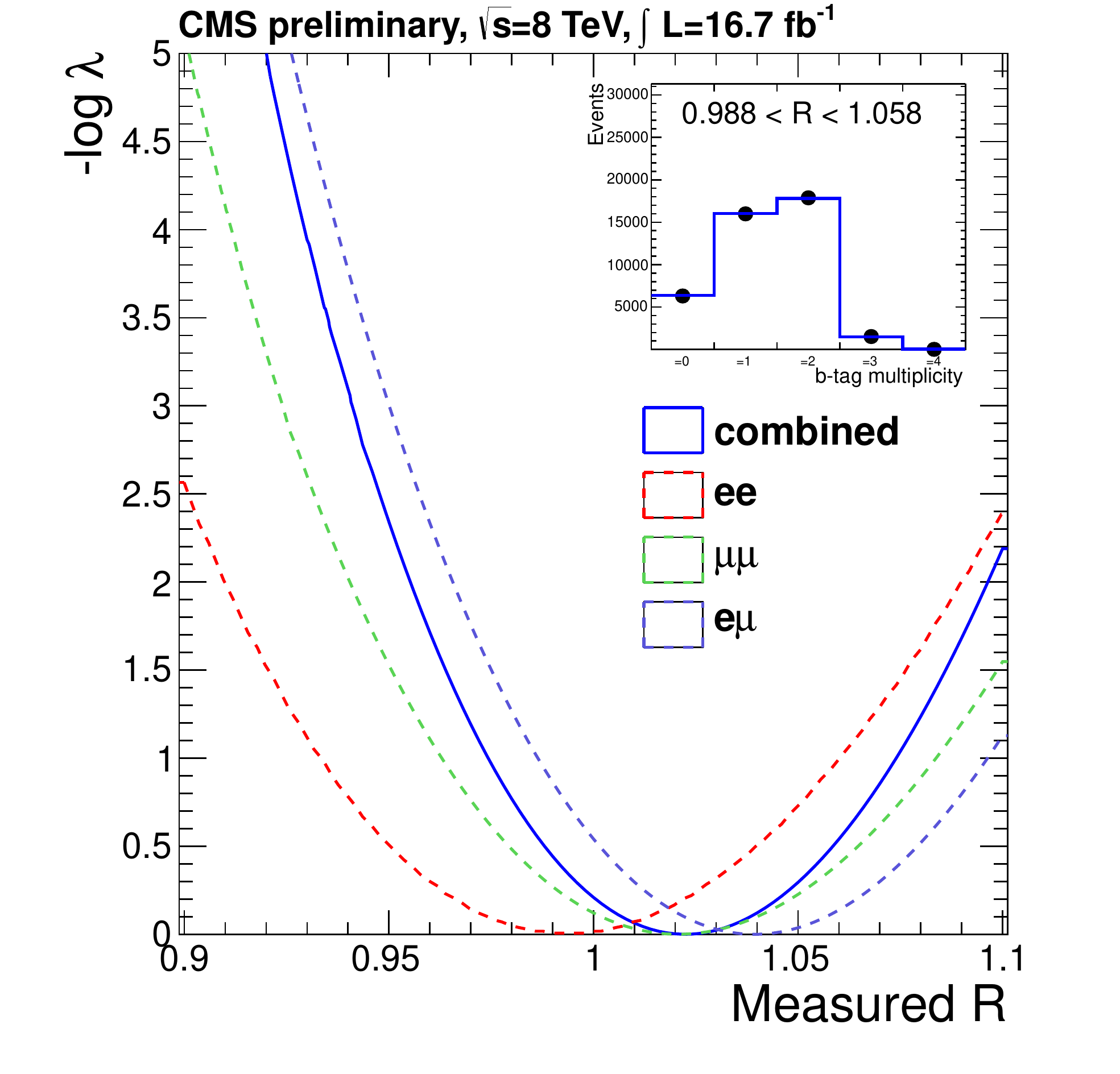}}
\end{minipage}
\hfill
\begin{minipage}{0.57\linewidth}
\centerline{\includegraphics[width=0.9\linewidth]{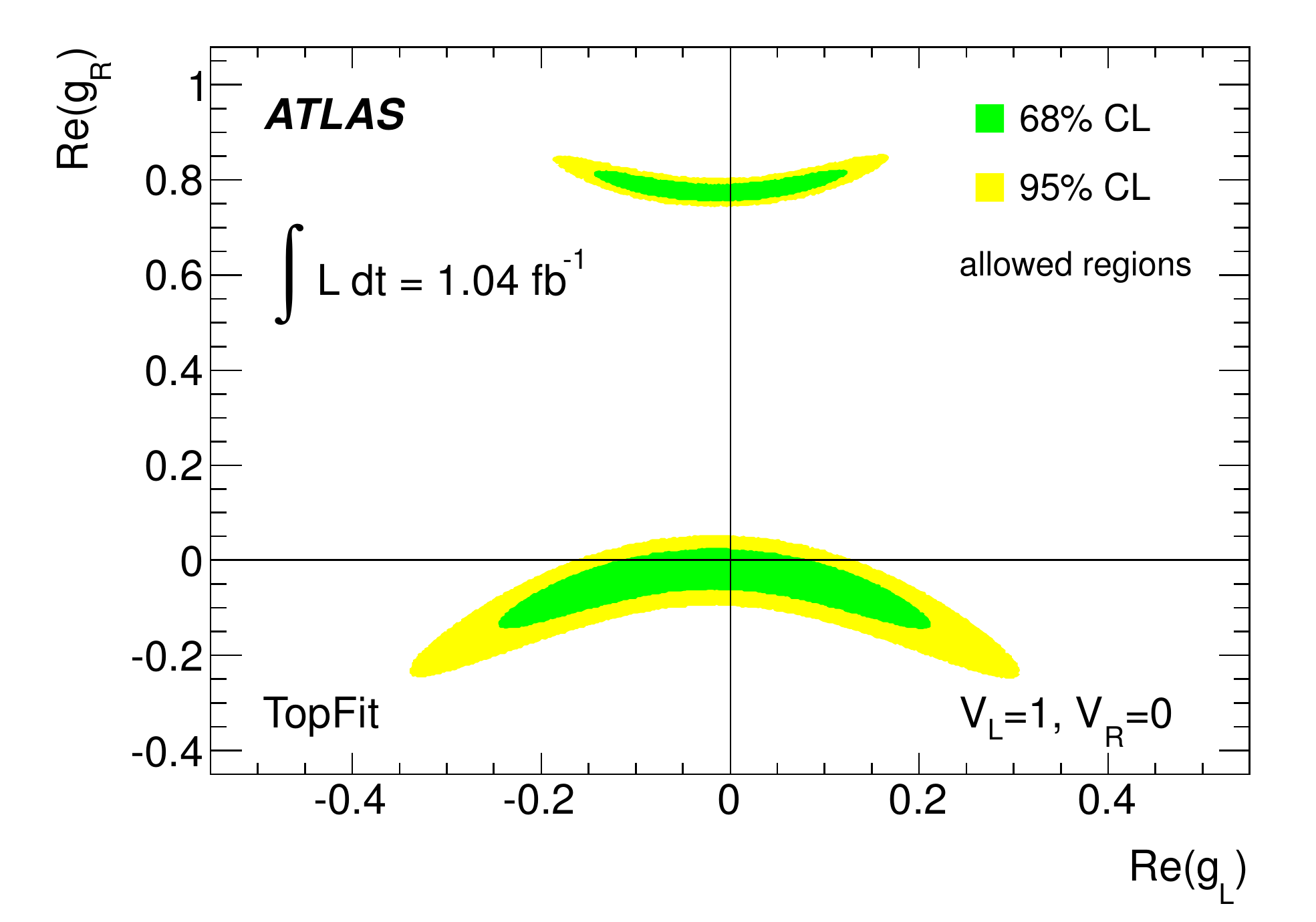}}
\end{minipage}
\hfill
\caption{Variation of the profile likelihood used to estimate $R$ from data for $\mu\mu$, $ee$, and $e\mu$
              dileptonic channels and for the one based on their combination (left).
              The insert shows the $b$-tag multiplicity distribution at the result of the combined fit.
              Allowed regions for $68\%$ and $95\%$ C.L. on the $tWb$ anomalous couplings $g_{L}$ and $g_{R}$
              from helicity fraction values measured by ATLAS (right).}
\label{fig:prop_R_helicity}
\end{figure}

The first one investigates possible deviations from the expected branching ratio $\mathcal{B}( t \ra Wb)$
that can be generically probed computing the ratio $R$ = $\mathcal{B}( t \ra Wb) / \mathcal{B}( t \ra Wq)$.
The value of $|V_{tb}|$ can be measured, validating the SM, or probing anomalies that could be hint
of BSM physics (e.g. a 4$^{th}$ generation of quarks or the existence of a light charged Higgs boson).
CMS has recently published a measurement of $R$ based on $L$ = 16.7 fb$^{-1}$ of 8 TeV
data \cite{prop:cms_R}.
The analysis, using dileptonically decaying top pairs, exploits a model that relates the expected
b-tagged jet multiplicity with $R$. The latter is extracted from data, profiting from the model,
by means of a binned profile likelihood fit on the inclusive multiplicity of jets and the one of 
b-tagged jets.
Results of the fit are depicted in Figure \ref{fig:prop_R_helicity} (left). The measured value of 
$R$ = $1.023^{+0.0036}_{-0.0034}$ is in good agreement with the SM and represent the most precise
world estimation at present.
Assuming the unitarity of the CKM matrix, it is possible to extract $|V_{tb}|$ = $1.011^{+0.0018}_{-0.0017}$
and constrain to $|V_{tb}| > 0.972$ ($95\%$ C.L.).

The second topic covered is the measurement of fraction of left-handed ($F_{L}$), right-handed
($F_{R}$) and longitudinally ($F_{0}$) polarised W bosons coming from top decays.
These quantities are well known in the SM and deviation from theoretically expected values are of
physical  interest as they can probe the $V$-$A$ nature of the SM model and measure anomalous coupling
of the $tWb$ vertex.
The helicity fractions are experimentally measurable by computing the $\theta^*$ angle, defined as the angle 
between the direction of the leptonic charge decay product of the $W$ boson and the inverse direction of the $b$ 
quark that arises from the $t$ decay, computed in the $W$ rest frame.
The latter is, in fact, related to the helicity fractions by:

\begin{equation}
\frac{1}{\Gamma}\frac{d\Gamma}{dcos(\theta^*)} = \frac{3}{8}(1-cos(\theta^*))^2F_{L} + \frac{3}{8}(1+cos(\theta^*))^2F_{R} + \frac{3}{4}sin^2(\theta^*)F_{0} .
\label{eq:costh_fractions}
\end{equation}

The ATLAS collaboration has experimentally performed such a measurement using the lepton+jets
and dileptonic top decay channels on a sample of  $L$ = 1.04 fb$^{-1}$ of 7 TeV data \cite{prop:atlas_heliticy}.
In the study the helicity fractions are extracted from the $cos(\theta^*)$ distribution by means of a binned
likelihood fit that compares data with templates generated using different helicity fractions in MC.
The same quantities are also measured using angular asymmetries defined as 
$A_{\pm} = (N(cos\theta^*>z) - N(cos\theta^*<z))/(N(cos\theta^*>z) - N(cos\theta^*<z))$ with $z = \pm (1-2^{2/3})$.
This method has the advantage of being less prone to some systematics.
Results from the two analyses are combined to get the final estimation of the helicity fractions.
CMS has recently performed a similar measurements in the dileptonic channel \cite{prop:cms_helicity_dilep}
($L$ = 1.04 fb$^{-1}$ collected at 7 TeV) and in single-top topology events \cite{prop:cms_helicity_singlet} 
($L$ = 1.14 fb$^{-1}$ of 7 TeV and $L$ = 5.3 fb$^{-1}$ of 8 TeV data).
In both studies, the $cos(\theta^*)$ distribution at reconstructed level is used to extract the values of
 $F_{L/R/0}$ that are computed by means of a fit based on a reweighing technique.
Results for all the analyses are summarised in Table 3. Good agreement with SM
predictions is observed. 
The presence of new physics can be parametrized in terms of an effective lagrangian that can be 
expressed as a function of anomalous couplings \cite{prop:theory_heliticy} corresponding, in the SM
at tree level, to $V_{L}$=$V_{tb}$,  $V_{R}$=0, $g_{R}$=0, $g_{L}$=0. Their relation
with W polarisation helicities is described in \cite{prop:theory_couplings}. 
Figure \ref{fig:prop_R_helicity} shows the limits imposed by the ATLAS helicity measurements
on the real component of $g_{R}$, $g_{L}$ coupling at the 68\% and 95\% confidence level.
Results are compatible with absence of anomalous $tWb$ couplings.

\begin{table}[t]
\label{tab:helicity}
\vspace{0.1cm}
\begin{center}
\begin{tabular}{lccc}
\hline
Experiment  & $F_{0}$ & $F_{L}$ & $F_{R}$ \\
\hline
\hline
\noalign{\vskip 1mm}
ATLAS                    &   0.67 $\pm$ 0.03  $\pm$ 0.06        &  0.32 $\pm$ 0.02  $\pm$ 0.03        &  0.01 $\pm$ 0.01  $\pm$ 0.04       \\
CMS$_{(dilep.)}$     &  0.698 $\pm$ 0.057  $\pm$ 0.148  &  0.288 $\pm$ 0.035  $\pm$ 0.083  &  0.014 $\pm$ 0.027  $\pm$ 0.087 \\
CMS$_{(single\ t)}$ &  0.713 $\pm$ 0.114  $\pm$ 0.023  &  0.293 $\pm$ 0.069  $\pm$ 0.030  &  -0.006 $\pm$ 0.057  $\pm$ 0.027 \\
\noalign{\vskip 1mm}
SM$_{NLO}$           & 0.687 $\pm$ 0.005                         & 0.311 $\pm$ 0.005                         & 0.0017 $\pm$ 0.0001 \\       
\hline
\end{tabular}
\end{center}
\caption{Summary of results on computation of helicity fractions and SM predictions.}
\end{table}

%%%%%%%%%%%%%%%%%%%%%%%%%%%%%%%%%%%%%%%%%%%%%%%%%%
% CONCLUSIONS
%%%%%%%%%%%%%%%%%%%%%%%%%%%%%%%%%%%%%%%%%%%%%%%%%%

\section{Conclusions}
The LHC collider has operated remarkably well allowing the ATLAS and CMS collaborations to quickly
collect an impressive set of results on the study of the top quark. Many measurements have 
already reached very good precision and, overall, excellent agreement with the standard model has
been encountered so far.

In this article an overview of the many aspects of this field was covered, going from 
production of top pairs and single top, to the study of the particle's characteristics and 
of its decay properties.
The large statistics of collected tops allowed to perform refined differential measurements and 
has implied that systematical uncertainties often dominate the incertitude of present
results.
The challenge for the future is thus to decrease systematics. More precise measurement on the
top sector are in fact crucial for new physics searches, where the top constitutes an important 
background or is part of the signal signature.

%%%%%%%%%%%%%%%%%%%%%%%%%%%%%%%%%%%%%%%%%%%%%%%%%%
% BIBLIOGRAPHY
%%%%%%%%%%%%%%%%%%%%%%%%%%%%%%%%%%%%%%%%%%%%%%%%%%

\end{document}